# Proof of Bell's Inequalities for Generalized Hidden Variables Models

### Angel G. Valdenebro

Lucent Technologies
Valderrobres, 10. 28022 Madrid. Spain
E-mail: avgonzalez@lucent.com

**Abstract**
Several authors have recently claimed that Bell's inequalities (BI) do not apply to certain types of generalized local hidden variables (HV) models. These claims are rejected, by means of a proof of BI valid for a very broad class of local HV models (deterministic or stochastic, with or without memory, time dependent or time independent, setting dependent or setting independent, and even allowing classical communication between the wings of the experiment). The precise meaning of the locality requirement is clarified during the proof. The exact roles of the two assumptions that amount to Bell's locality (parameter independence and outcome independence) are explicitly shown, using each one of them in a different step of the proof.

## 1   Introduction

Proofs of Bell's inequalities (BI) are anything but scarce in the literature (references [1-7] are representative examples), so I will try to explain, first of all, what is the interest of a new, more general proof. BI can be proved for hidden variables (HV) models that satisfy some assumptions. Locality is the most important of these assumptions, but not the only one [8]. The theorem of Bell states that, because quantum mechanics predictions violate the inequalities (predictions that are confirmed by many experimental results), these types of HV models can be discarded as models of reality, and as realistic theories underlying quantum mechanics. The more general a proof of BI is, the bigger class of HV models it discards. Or, to put it in positive terms, the more general a proof of BI is, the more it delimits the HV models that can underlie quantum mechanics and constitute valid models for reality.

Recently, several authors have proposed generalized HV models not considered in previous proofs of BI, claiming that these models can evade Bell's conclusions, reproducing the predictions of quantum mechanics while respecting all the other assumptions of BI, and in particular, while being local. One of the objectives of the present work is to extend the proof of BI to these models, rejecting the corresponding claims.

A first generalization of the concept of HV is the inclusion of memory effects [9]. The predictions that these models make for the result of a certain measurement may depend on the previous sequence of measurements performed on the system and the outcomes obtained. Several articles [10,11] have shown that memory effects do not invalidate Bell's conclusions, but not by presenting a proof of BI compatible with these new type of HV, but by using elaborated statistical arguments.

In HV theories the state of the observed object is represented by means of an ensemble of microstates labeled by HV. Hess and Philipp [12-14] propose another generalization of HV models where also the measuring devices are described by microstates and HV. Besides, they argue that the HV of the observed object and of the experimental devices may depend on time, and may be correlated by means of synchronized internal clocks. Although Hess and Philipp do not consider the debate closed, the model that they present as an example of their theory has been criticized for being non-local [7,15-17]. Anyway, they have raised an important issue: previously known proofs of BI based on HV do not work for time-correlated setting-dependent HV.

The proof presented here works for HV that can be deterministic or stochastic, with or without memory, time independent or time dependent, setting independent or setting dependent. In fact, the proof works for a very broad class of HV models that allow the microstate of the observed object or of the measuring devices or both, to depend not only on history (as in HV models with memory) or on time (as in the model of Hess and Philipp), but also on information exchanged among the devices and the object by means of classical communication. The only limitation is that the choice of measurements to be performed at each trial could be made after the exchange of information is concluded. Or, in other words, the only information that can not be exchanged before a trial is finished is the choice of settings of the experimental devices for this trial. The presence of classical communication in a Bell type experiment has been analyzed by Mermin in a recent note [18], although he has not proved BI in general, but justified the conclusions of the theorem of Bell from a particular geometry and sort of experiment.

To analyze the relationship between the locality requirement of BI and relativistic locality in detail, it is convenient [19,20,21] to consider Bell's locality as a combination of two other assumptions: parameter independence and outcome independence. This is the first proof of BI where the exact role of parameter independence and outcome independence assumptions is explicitly shown, because each one of them is used in a different step of the proof. Consequently, an additional advantage of this proof is that it contributes to clarify the meaning of parameter independence and outcome independence, and their need for BI.

## 2  Terminology

Let us clarify notation and terminology. We are dealing with a typical Bell type experiment. A couple of entangled particles (photons for example) are emitted in different directions from a central source. For each "trial" (i.e. for each couple of emitted particles) Alice, at the "A" wing of the experiment, can measure one of the observables $A$ or $A'$ on the particle she receives, while Bob, at the "B" wing of the experiment, spatially separated from Alice, can measure $B$ or $B'$ on his particle. A "run" of the experiment is a long sequence of trials, from which the correlations: $\langle AB \rangle$, $\langle A'B \rangle$, $\langle AB' \rangle$ and $\langle A'B' \rangle$ can be obtained. To avoid problems with the "detection loophole" we will suppose either that all the intended measurements are successful, or that the reasoning below refers always to (and the assumptions apply to) "successful" trials. The four observables are taken to be dichotomic (i.e. with only two possible outcomes: 1 and –1) or, at least, bounded by ±1. Under some assumptions, the values of the correlations are constrained by certain limits called BI. In particular, we will prove the CHSH [3] version of BI:

$$\left|\langle AB'\rangle + \langle B'A'\rangle + \langle A'B\rangle - \langle BA\rangle\right| \leq 2. \tag{1}$$

## 3 The proof

Our first assumption is that we have a HV model that reproduces the experimental results. The meaning of this assumption is the following. The HV model can provide possible results for each trial of the experiment. Given an instant of time, and a choice of experimental settings (a choice of observables to be measured), the model provides a distribution of compatible microstates for the system. By system we understand the whole experimental setup, so the model may include setting-dependent HV describing not only the microstate of the measured object but also the microstate of the experimental devices. We can randomly draw a certain microstate from that distribution. Now, given the microstate of the system, the model must provide probability distributions for the possible outcomes of the measurements (if necessary, time dependencies and information exchanged by the devices may be taken into account when calculating those distributions). Drawing outcomes from these probability distributions, the model provides possible results for a certain trial of the experiment. Given a sequence of time instants, we can repeat the previous procedure for each instant of time, and the model will provide potential results for a complete run of the experiment. If the model includes "memory" effects, the previous sequence of trials will be taken into account at each trial. That the HV model reproduces experimental results means that, for any long enough run, the correlations calculated from the potential results obtained from the model, must coincide, within experimental error, with the values obtained in a real experiment.

Let us begin the proof by building a table of potential outcomes as follows. At each instant of time we have four different possible settings, that we will denote by numbers from 1 to 4, corresponding to the four possible pairs of observables being measured: $(A, B)$, $(A', B)$, $(A, B')$, and $(A', B')$ respectively. We have to draw a microstate for each pair of measurements. If the model were not setting-dependent the microstate would refer only to the observed object and could be the same for all the four different experimental settings, but we are allowing setting dependencies, so we can consider four different microstates if necessary. From each microstate we draw possible outcomes for the corresponding measurements. We can write the obtained values in eight columns, labeling each column with a letter that indicates the measured observable and a subindex that indicates the setting: $\{A_1, B_1, A_2, B'_2, A'_3, B_3, A'_4, B'_4\}$. Now we will record which of the four pairs of observables is actually measured in the experiment we are reproducing, and annotate the corresponding number (from 1 to 4) in an additional column that we will label as $S$ (for setting). We can repeat the procedure for each trial in the experiment, adding one row to the table each time. As we explained above, memory effects, time dependencies, setting dependencies and classical communication can be taken into account when filling up the table.

To calculate any correlation (for example $\langle AB\rangle$) from the table, we have to consider only those rows corresponding to the correct setting (rows with $S = 1$ in the example), and only the values on the correct columns ($A_1$ and $B_1$ in the example).

Our second assumption, usually called "no conspiracy", states that, for the kind of experiment whose results we are reproducing, the choice of the experimental setting ($S$) for each trial is made in a way stochastically independent from the microstate of the system and from the potential outcomes calculated from it. Some authors relate this assumption to the "free will" of the experimenters. But we do not need philosophical discussions about determinism here. Even radical determinists could agree that there are methods (coin tosses, random number generators, etc.) to choose the experimental setting (i.e. the value in column $S$) so that it is stochastically independent from the values in the columns $\{A_1, B_1, A_2, B'_2, A'_3, B_3, A'_4, B'_4\}$ for the same row. Technically this means that if we decompose the table in subtables formed selecting the rows that contain the same set of values for $\{A_1, B_1, A_2, B'_2, A'_3, B_3, A'_4, B'_4\}$, the statistical distribution of values of $S$ is the same in all the subtables. It cannot be other way because each $S$ has been chosen independently of the values of the other columns.

Stochastic independence is reciprocal. If $S$ is stochastically independent of $\{A_1, B_1, A_2, B'_2, A'_3, B_3, A'_4, B'_4\}$ then $\{A_1, B_1, A_2, B'_2, A'_3, B_3, A'_4, B'_4\}$ is also stochastically independent of $S$. Therefore, if we form subtables with the same value of $S$, the statistical distribution of values of $\{A_1, B_1, A_2, B'_2, A'_3, B_3, A'_4, B'_4\}$ is the same in all the subtables. It does not matter that in a HV with memory the outcomes may depend on the settings of previous trials. It does not matter the possible existence of setting-dependencies in the HV. The key is that $S$ is selected randomly for each trial.

Therefore, to calculate a correlation (such as $\langle AB \rangle$) it does not matter if we use only the rows pertaining to the corresponding setting (rows with $S=1$) or we use all the rows of the table, because the joint distribution of the variables ($A_1$ and $B_1$) is the same in both cases.

The third assumption is that the HV model features parameter independence. Parameter independence states that, for a given trial, the probability of an outcome of an observation on the "A" wing is independent of the experimental setting on the "B" wing. This means that for each trial, the probability distributions from which $A_1$ and $A_2$ are drawn are identical to each other. If the probability distributions are the same in each trial, after a large enough number of trials the statistical distribution of $A_1$ and $A_2$ will be the same. Therefore, if we consider the whole table, each possible value appears the same number of times in column $A_1$ than in column $A_2$. We can reorder the data in columns $A_2$ and $B'_2$ (values in both columns together) so that the values of $A_1$ and $A_2$ coincide in each row. Because the value of an average does not depend on the order of the summands, this reordering does not change the value of $\langle AB' \rangle$. Once the columns $A_1$ and $A_2$ are identical we will use the same label for both of them and denote our table by $\{A, B_1, A, B'_2, A'_3, B_3, A'_4, B'_4\}$.

We can apply exactly the same reasoning to columns $B_1$ and $B_3$. Parameter independence implies that their statistical distributions are the same. We can reorder the data in columns $A'_3$ and $B_3$ until columns $B_1$ and $B_3$ coincide, and write $\{A, B, A, B'_2, A'_3, B, A'_4, B'_4\}$, without affecting $\langle A'B \rangle$. Repeating once more the procedure, we reorder the data in columns $A'_4$ and $B'_4$ until columns $B'_2$ and $B'_4$ are identical, transforming our table into $\{A, B, A, B', A'_3, B, A'_4, B'\}$ without affecting $\langle A'B' \rangle$.

Finally, we would like to do something similar to columns $A'_3$ and $A'_4$. But $A'_3$ and $A'_4$ belong to pairs of columns that have been reordered already, so we have to be very careful not to destroy what we have done until now. We need one final assumption to complete our task: the HV model features outcome independence. Outcome independence (predictive completeness in Ballentine and Jarrett's terminology [21]) means that the outcome of an observation at the "B" wing provides no information relevant to predicting the outcome of an observation at the "A" wing that is not already contained in the microstate and time instant specification. In other words, for a given trial (i.e. for a given time and microstate), the probability of an outcome of an observation on the "A" wing is independent of the outcome of the observation on the "B" wing.

We already knew that, due to parameter independence, the statistical distributions of $A'_3$ and $A'_4$ are equal to each other when we consider all the rows of the table. Are the distributions of $A'_3$ and $A'_4$ equal to each other also in a large enough subtable extracted from the table? It depends on the criterion used to select the rows that form the subtable. If the criterion is based only on time and/or microstate of the system, the equality of the statistical distributions holds, because parameter independence implies that the probability distributions from which we draw $A'_3$ and $A'_4$ are identical *for each instant of time and microstate*. Outcome independence tells us that the value of $B'_4$ (that we are now calling simply $B'$) does not provide any information relevant to predicting the value of $A'_4$ (nor of $A'_3$) that is not already provided by the time and microstate of the system. This means that selecting the rows that form a subtable by the value of $B'$ cannot have any effect on the statistical distributions of $A'_3$ and $A'_4$ (in particular breaking their equality) that could not be obtained selecting the rows by time and/or microstate. Therefore, the statistical distributions of $A'_3$ and $A'_4$ are equal to each other in each subtable formed by all the rows with the same value of $B'$. We can reorder the values of $A'_4$ *within each of these subtables* until the columns $A'_3$ and $A'_4$ coincide. Because we restrict ourselves to reorder the values of $A'_4$ in rows with the same value of $B'$, we are not affecting the value of $\langle A'B' \rangle$ and not destroying any previous reordering.

We have finally transformed our table into $\{A, B, A, B', A', B, A', B'\}$, or removing repeated columns, into $\{A, B, B', A'\}$. Thanks to our assumptions (no conspiracy, parameter independence and outcome independence), the correlations that we calculate from this simplified table taking into account all the rows, are equal to the correlations calculated from the original table $\{A_1, B_1, A_2, B'_2, A'_3, B_3, A'_4, B'_4\}$ taking into account only the rows pertaining to the corresponding setting. In other words, we have built a joint distribution for variables $A$, $A'$, $B$ and $B'$, and we can calculate all the correlations as simple averages over this joint distribution.

Once we have arrived to this result, it is well known that BI can be easily derived, so we will only sketch the derivation

$$\left|\langle AB'\rangle + \langle B'A'\rangle + \langle A'B\rangle - \langle BA\rangle\right|$$
$$= \left|\langle AB' + B'A' + A'B - BA\rangle\right|$$
$$\leq \langle |AB' + B'A' + A'B - BA|\rangle$$
$$= \langle |A(B' - B) + A'(B' + B)|\rangle$$
$$\leq \langle |A||B' - B| + |A'||B' + B|\rangle \quad (2)$$
$$\leq \langle \max(|A|, |A'|)(|B' - B| + |B' + B|)\rangle$$
$$\leq \langle \max(|A|, |A'|) \cdot 2\max(|B|, |B'|)\rangle$$
$$= 2\langle \max(|A|, |A'|) \max(|B|, |B'|)\rangle$$
$$\leq 2,$$

where we have used the following facts: (i) a sum of averages is equal to the average of the sum, (ii) the absolute value of an average is less than or equal to the average of absolute values, (iii) the absolute value of a sum is less than or equal to the sum of absolute values, and (iv) the absolute values of all the variables are bounded by 1.

Note that we have used the phrases "long enough run", "large enough number of trials", and "large enough subtable" during the proof. In any real run, composed by a finite number of trials, the distributions of the results could deviate from the ideal distributions that we have considered above. However, the law of large numbers assures us that these deviations become negligible as the number of trials increases. In other words, what we have proved is that the expression $\left|\langle AB'\rangle + \langle B'A'\rangle + \langle A'B\rangle - \langle BA\rangle\right|$ is bounded by a quantity that tends to 2 as the number of trials tends to infinity.

## 4 Conclusion

HV models featuring parameter independence and outcome independence predict results in accordance with BI for experiments where the choice of observables to be measured at each successful trial is independent from the value of the HV. This conclusion is valid for a very general class of HV, including those where the determination of (probabilities of) the possible outcomes for a certain trial may be influenced by: (i) settings and outcomes of previous trials, (ii) the microstate of the experimental devices, (iii) time, and (iv) any kind of information exchanged classically among the experimental devices and the observed object, before the choice of settings is made.

## 5 Acknowledgments

I am indebted to Karl Hess and Walter Philipp for helpful criticisms of an earlier version of the manuscript.